\documentclass[11pt]{article}
\usepackage{moriond,epsfig}

\def\Journal#1#2#3#4{{#1} {\bf #2}, #3 (#4)}

\def\NPB{{\em Nucl. Phys.} B}

\def\be{\begin{equation}}
\def\ee{\end{equation}}
\def\bea{\begin{eqnarray}}
\def\eea{\end{eqnarray}}

\begin{document}
\vspace*{4cm}
\title{GENERALIZED CHAOTIC INFLATION}

\author{ F.J. CAO }

\address{(a) LPTHE, Universit\'e Pierre et Marie Curie (Paris VI) \\ et Denis Diderot (Paris VII), Tour 16, 1er. \'etage, \\ 4, Place Jussieu, 75252 Paris cedex 05, France\\
(b) Complexo Interdisciplinar, Universidade de Lisboa, \\ Avda. Prof. Gama Pinto, 2, 1649-003 Lisbon, Portugal}

\maketitle\abstracts{ We investigate inflation driven by the evolution of highly excited {\em quantum states} within the framework of out of equilibrium field dynamics.
The self-consistent evolution of these quantum states and the scale
factor is studied analytically and numerically. It is shown that the
time evolution of these  quantum states lead to two consecutive stages
of inflation under conditions that are the quantum analogue of
slow-roll. The evolution of the scale factor during the first stage
has new features that are characteristic of the quantum state.  During
this initial stage the quantum fluctuations in the highly excited band
build up an effective homogeneous condensate with a non-perturbatively
large amplitude as a consequence of the large number of quanta.
The second stage of inflation is similar to the usual classical
chaotic scenario
but driven by this effective condensate. The excited quantum
modes are already superhorizon in the first stage and do not affect the
power spectrum of scalar perturbations.  Thus, this novel scenario provides a field theoretical justification for chaotic scenarios driven by a classical homogeneous scalar field of large amplitude.}

\section{Introduction}

Inflation is a stage of accelerated expansion in the very early Universe. The present observations makes inflationary cosmology be the leading theoretical framework to explain the homogeneity, isotropy and flatness of the Universe, and the observed features of the cosmic microwave background.

There are very many different models for inflation motivated by particle physics and most if not all of them invoke one or several scalar fields, the inflaton(s), whose dynamical evolution in a scalar potential leads to an inflationary epoch. These inflaton fields are scalar fields that attempt to provide an effective description for the fields in the grand unified theories.

Most treatments of inflation study the evolution of the inflaton as a {\em homogeneous classical scalar} field. The quantum field theory interpretation is that this classical homogeneous field configuration is the expectation value of a quantum field operator in a translational invariant quantum state. In this treatments, while the evolution of this coherent field configuration is studied via classical equations of motion, quantum fluctuations of the scalar field around this expectation value are treated perturbatively and are interpreted as the  seeds for scalar density perturbations of the metric\cite{infbooks}.

An important class of inflationary models produces inflation starting from highly excited states that rolls down the potential (for example: chaotic inflation). This models are usually studied in a classical framework. 
In order to have a long inflationary period, it is necessary that the field rolls down very slowly. For this models various conditions have been obtained that are different realizations of what we will call the {\em classical slow roll condition},
\be
\dot \eta^2 \ll |m^2| \eta^2,
\ee
this condition guarantees that there is inflation and that it last long.
($ \eta $ is the amplitude of the classical scalar field, $ m $ its mass, and the dot denotes time derivative).

The purpose of this article is to provide a consistent {\em quantum treatment} of models whose classical counterparts are these ``large field" models.

\section{Initial conditions and equations of motion}

We treat the inflaton as a full quantum field, and we study its dynamics in a classical space-time metric (consistently with inflation at a scale well below the Planck energy density). The dynamics of the classical space-time metric is determined by the Einstein equations  with a source term given by the expectation value of the energy momentum tensor of the quantum inflaton field ($ G_{\mu\nu} = 8\pi M_{Pl}^{-2}\langle T_{\mu\nu} \rangle $).
Hence  we solve  {\em self-consistently} the coupled evolution equations for the classical metric and the quantum inflaton field.

We assume that the universe is homogeneous, isotropic and spatially flat, thus it is  described by the metric,
\begin{equation}\label{metric}
ds^2 = dt^2 - a^2(t)\; d\vec{x}^2\; .
\end{equation}
We consider an inflaton model with the matter action given by
\begin{equation}
S [\vec\Phi] =  \int d^4x\; {\cal L}_m = \int d^4x \;
a^3(t)\left[\frac{1}{2} \, \dot{\vec{\Phi}}^2(x)-\frac{1}{2} \,
\frac{(\vec{\nabla}\vec{\Phi}(x))^2}{a^2(t)}-V(\vec{\Phi}(x))\right]\; ,
\label{action}
\end{equation}
\begin{equation}
V(\vec{\Phi})  =  \frac{m^2}2\; \vec{\Phi}^2 +
\frac{\lambda}{8N}\left(\vec{\Phi}^2\right)^2 \;, \label{potential}
\end{equation}
where $\vec{\Phi}(x) $ is an $N$-component scalar inflaton field. Here, we study the case $ m^2 > 0 $.

We investigate the possibility of inflation driven by the evolution of highly excited {\em quantum} state with large energy density. This implies the need a non-perturbative treatment of the evolution of the quantum state and therefore we use the large $N$ limit method.

In order to obtain the evolution equations it is convenient to decompose the field as follows
\be
\vec\Phi(\vec x,t) = \left(\sqrt{N}\,\eta(t)+\chi(\vec x,t),\;\vec\pi(\vec x,t)\right)
\ee
where $\langle\vec\Phi\rangle = (\sqrt{N}\,\eta(t),\;\vec 0)$, and
\be
\vec{\pi}(\vec x,t) = \int \frac{d^3 k}{\sqrt{2}(2\pi)^3} \left[
\vec{a}_{k} \; f_{k}(t) \;e^{i\vec k \cdot \vec x}+
\vec{a}^{\dagger}_{k} \; f^*_{k}(t) \; e^{-i\vec k \cdot \vec
x}\right] 
\ee
here, $f_k$ are the modes of the quantum fluctuations (note that they can be large in this non-perturbative framework).

The evolution equations for the field in the large $N$ limit are \footnote{the contributions of $ \chi $ are subleading},
\begin{eqnarray}
&&\ddot\eta + 3\,H\,\dot\eta + {\cal M}^2\,\eta = 0 \cr\cr &&\ddot
f_k +3\,H\,\dot f_k + \left(\frac{k^2}{a^2} + {\cal
M}^2\right)\,f_k = 0 \quad\quad
\mbox{with }\quad {\cal M}^2 = m^2 + \frac{\lambda}{2}\,\eta^2 +
\frac{\lambda}{2}\,\int_R{\frac{d^3k}{2(2\pi)^3}\,|f_k|^2},
\end{eqnarray}
and for the scale factor ($ H = \dot a / a $),
\be
H^2 = \frac{8\pi}{3\,M_{Pl}^2} \; \epsilon\;;
\quad\quad\frac{\epsilon}{N} = \frac12 \, \dot\eta^2 + \frac{{\cal M}^4 -
m^4}{4} +\frac14 \int_R \frac{d^3k}{(2\pi)^3} \left(|\dot f_k|^2 +
\frac{k^2}{a^2}|f_k|^2\right)\;.
\ee
where $ \epsilon = \langle T^{00} \rangle $ is the energy density.
(The subindex $ R $ denotes the renormalized expressions for these integrals, the explicit expressions for the subtractions can be found in Ref. 2.)
The equations for the expectation value and for the field modes are analogous to damped oscillator equations, and the inflationary period would correspond to the overdamped regime of this damped oscillators.

We focus on the possibility of inflation through the dynamical {\em
quantum} evolution of  a highly excited initial state with large energy
density. 
It can be shown that the initial conditions for a general pure state can be parameterized as follows\cite{tsuinf,tsunos},
\be 
f_{ k}(0) =  \frac{1}{\sqrt{\Omega_{k}}} \;; \quad \quad
\dot f_{ k}(0) =  - [\omega_k(0)\,\delta_{ k} + H(0)
+ i \Omega_{ k}] \; f_{ k}(0) \label{inicondcomo} \; .
\ee
where $ \Omega_{k} $ and $ \delta_{k} $ set the initial conditions. They  fix the energy contained in each mode and the coherence between them.

\section{Generalized chaotic inflation}

In this framework we have found that the following \emph{generalized slow roll condition}
\be
\dot\eta^2+\int_R{\frac{d^3k}{2(2\pi)^3}\,|\dot f_k|^2} \;\ll\; m^2
\left( \eta^2 + \int_R{\frac{d^3k}{2(2\pi)^3}\,|f_k|^2} \right)
\ee
guarantees there is inflation ($\ddot a > 0$) and that it last long.
(This condition includes the classical one $ \dot\eta^2 \ll m^2 \eta^2 $ as a particular case.)

This scenario presents \emph{two inflationary epochs}:
\begin{enumerate}
\item \emph{Fast decreasing Hubble parameter epoch}. During this epoch the term $ D \equiv \int \frac{d^3k}{(2\pi)^3} \frac{k^2}{a^2}\,|f_k|^2 $ has an important contribution to the energy density, and the dominant process is the redshift of the excitations ($ k/a \to 0 $). This epoch ends when this $ D $ contribution becomes negligible.
\item \emph{Quasi-De Sitter epoch}. The enormous redshift of the previous epoch assembles the quanta into a zero mode condensate, 
\be
\eta_{eff}(t) =
\sqrt{\eta^2+\int{\frac{d^3k}{2(2\pi)^3}\,|f_k|^2}}\;,
\ee
that verifies the classical equations of motion,
\be
\ddot \eta_{eff} + 3 \,H \,\dot\eta_{eff} + m^2 \, \eta_{eff}
  + \frac\lambda2\,\eta_{eff}^3 = 0\;.
\ee
Therefore, this period can be {\em effectively} described by a classical field.

\end{enumerate}

\begin{figure}
\begin{center}
\epsfig{figure=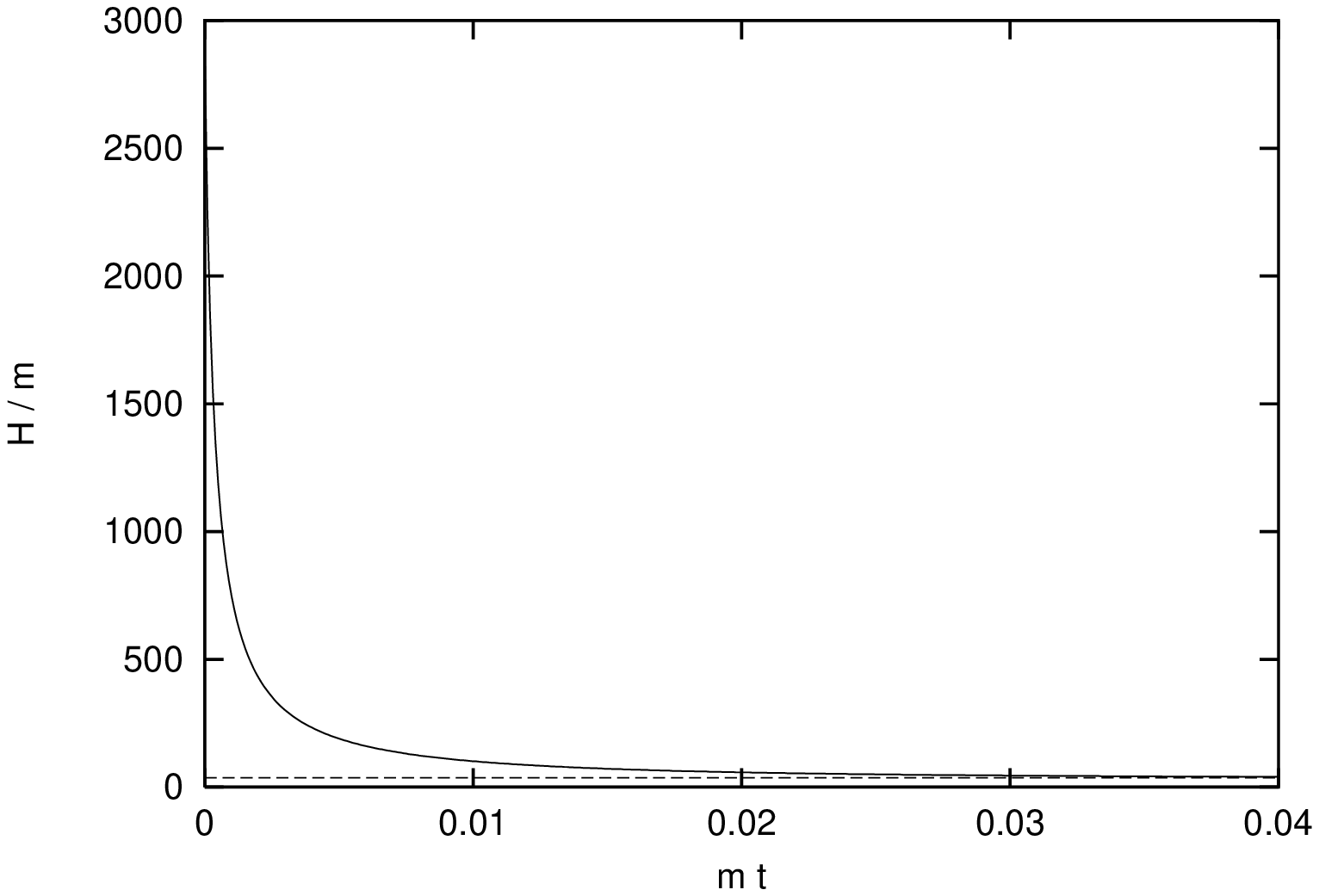,height=1.5in} \hspace{2cm}
\epsfig{figure=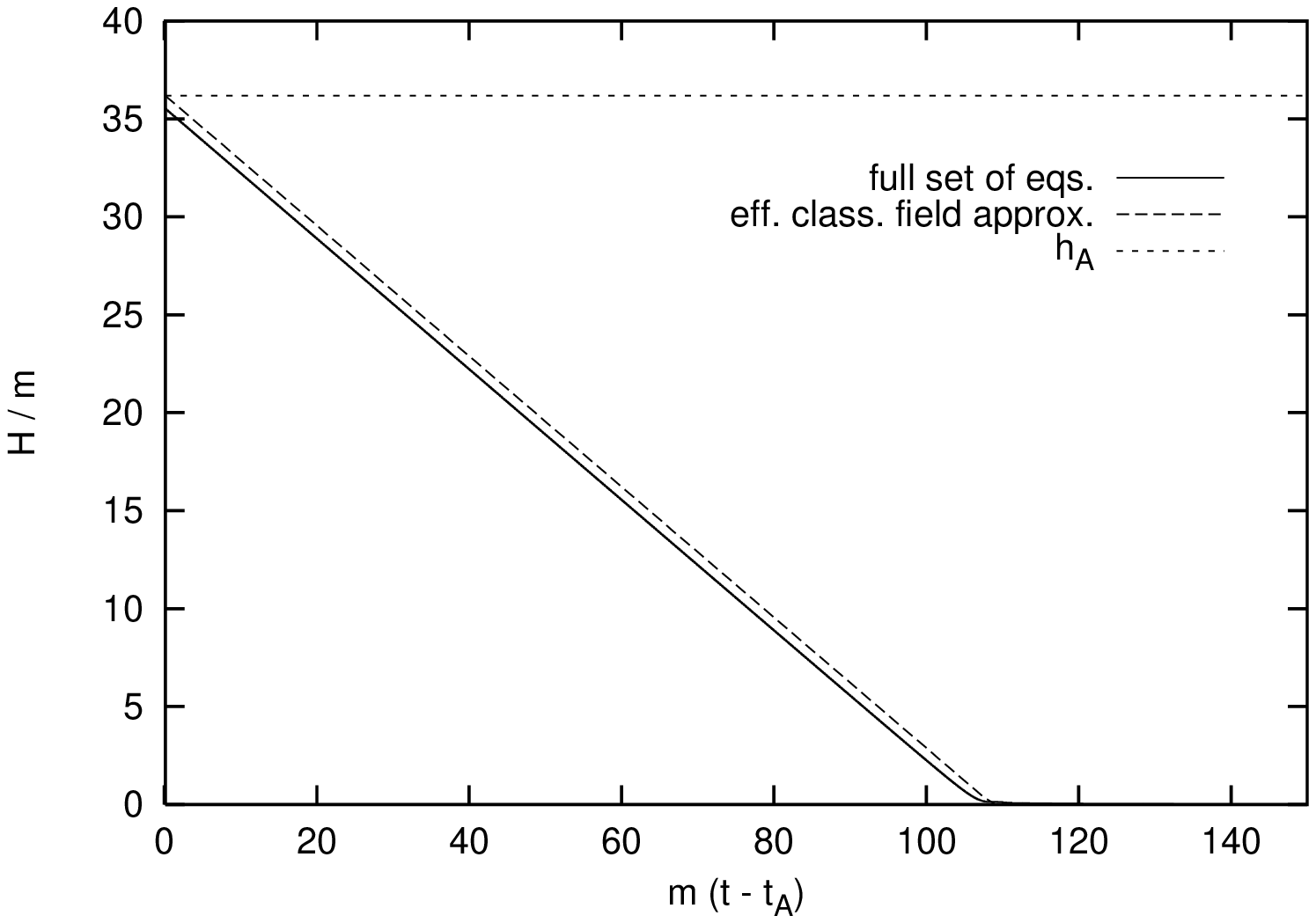,height=1.5in}
\end{center}
\caption{{\em Left}: H(t) during the inflationary epoch with fast decreasing of the Hubble parameter. For $ m = 10^{-4} M_{Planck} $, $ \lambda = 10^{-12} $ and $ N = 20 $. Initial conditions: $ \epsilon_0 = 10^{-2} M_{Pl}^4 $, and only a excited shell of quanta in $ q_0 = 80.0 $.
{\em Right}: H(t) in the subsequent quasi-De Sitter epoch for the same configuration (this period starts at $ t_A $ with the final value $ h_A $ of the previous period).}
\label{hearlyfig}
\end{figure}

The total number of e-folds ($ N_e \equiv log(a_{final}/a_{initial} $) decreases if the initial state had high momentum quanta excited (for constant initial energy). For example, if the initial energy is concentrated in a shell of momenta $ k_0 $ and the quadratic term in the potential (\ref{potential}) dominates; we have $ N_e \simeq \frac{4\pi}{M_{Pl}^2\,m^2}\; \frac{\epsilon_0}{\;1+(k_0/m)^2\;} $ (where the classical result is recovered at $ k_0 = 0 $).

\section{Comments and Conclusions}

In this article we study the chaotic inflation scenario giving a consistent non-perturbative quantum treatment to the inflation field, that allows to study the dynamics of highly excited quantum states.
We have shown that these highly excited quantum states lead to efficient inflation under conditions that are the quantum analog of the classical slow roll conditions. Therefore, it broads the class of initial states known to give rise to efficient inflation. (A further generalization to mixed states can also be done\cite{tsuinf}.)
It is also important to note that this generalized slow roll condition allows initial states that do not break the possible $ \vec\Phi \to -\vec\Phi $ symmetry of the potential.

Furthermore the classical chaotic inflation scenario emerge as an effective description of the second inflationary period.
Therefore, the generalized chaotic inflation provides a field theoretical justification for classical chaotic inflation.

\section*{Acknowledgments}

The results exposed here are part of the output of a collaboration with Hector J. de Vega and Daniel Boyanovsky.
This work has been supported by the MEC (Spain) and by the European Network HPRN-CT-2000-00158.

\end{document}